\documentclass[aps,pra,10pt,superscriptaddress]{revtex4-2}
\usepackage{amsmath}
\usepackage{tikz}
\usepackage{braket}
\newcommand{\ketbra}[2]{\left|#1\right\rangle\hskip-1mm\left\langle#2\right|}
\usepackage{orcidlink}
\usepackage{hyperref}
\usepackage{color}
\usepackage{ulem}
\usepackage[capitalise]{cleveref}
\crefname{equation}{Eq.}{Eqs.}
\crefname{figure}{Fig.}{Figs.}
\usepackage[authormarkuptext=name,commentmarkup=uwave]{changes}
\definechangesauthor[name={JRH},color=blue!70!black]{Jonte}

\begin{document}

\title{What meter interference can tell us about the statistics of weak measurements}


\author{Holger F. Hofmann\,\orcidlink{0000-0001-5649-9718}}
\email{hofmann@hiroshima-u.ac.jp}
\affiliation{Graduate School of Advanced Science and Engineering, Hiroshima University, Kagamiyama 1-3-1, Higashi Hiroshima 739-8530, Japan}

\author{Tomonori Matsushita\,\orcidlink{0000-0001-7713-5374}}
\affiliation{Graduate School of Advanced Science and Engineering, Hiroshima University, Kagamiyama 1-3-1, Higashi Hiroshima 739-8530, Japan}

\author{Masayoshi Isobe}
\affiliation{Graduate School of Advanced Science and Engineering, Hiroshima University, Kagamiyama 1-3-1, Higashi Hiroshima 739-8530, Japan}

\author{Jonte R. Hance\,\orcidlink{0000-0001-8587-7618}}
\email{jonte.hance@newcastle.ac.uk}
\affiliation{Quantum Group, School of Computing, Newcastle University, 1 Science Square, Newcastle upon Tyne, NE4 5TG, UK}

\begin{abstract}
It is widely assumed that the quantum fluctuations of the meter readout in a weak measurement make it impossible to identify the contributions originating from the individual values of the physical property observed in the measurement. Here, we show that a careful analysis of the quantum dynamics of the meter system allows us to identify a universal relation between quantum interference in the post-selection probability, and quantum interference in the statistics of the meter readout. The analysis reveals that quantum interference modifies the readout distribution of the meter in two ways, a diffusion term that identifies the appropriate operator ordering in the post-selected variance of the observed system property, and a wavefunction-dependent update of the initial meter statistics based on the effect of back action on the post-selection probability. Our results show that the statistical patterns described by meter interference provide important details about the physics of post-selection in weak measurements, allowing us to exclude the possibility of statistical artefacts in the experimental observation of weak values. 
\end{abstract}

\maketitle

\section{introduction}
\label{sec:introduction}

As an empirical science, physics should be grounded in experimentally observable facts. Unfortunately, quantum mechanics fails this criterion, by separating the description of the state of a system from the potential outcomes of a specific measurement on that system. In the Hilbert space formalism, operators assign values to physical properties by associating eigenvalues with an orthogonal basis system of eigenstates. This assignment fails to explain how measurements of different properties are related to each other, with the eigenstates of one property being composed of superpositions of eigenstates of the other. In general, the observation of a specific outcome potentially depends on quantum interference between eigenstates of any physical property that is not directly observed in that measurement. This interference between different eigenstates of an observable leaves it unclear how the probability of a measurement outcome described by a superposition of eigenstates relates to the possible observation of eigenvalues in a measurement described by projections onto these eigenstates. What does quantum mechanics tell us about the relation between eigenvalues and quantum interference between orthogonal eigenstates of the same physical property?

If one assumes that eigenvalues always represent the correct values of system observables, regardless of the initial system state and the measurement context, one encounters paradoxes such as quantum contextuality \cite{Spe60,KS68,Bud+22}, or violations of Bell inequalities \cite{Bel64,FC72,ADR82,Pan+98}. These paradoxes demonstrate that the relationship between different measurements cannot be consistently explained solely in terms of eigenvalues representing the outcomes of fully resolved measurements. It would seem natural to conclude from this that measurements performed at finite resolution should not be interpreted in terms of the eigenvalues observed at maximal resolution either, since the characteristic statistics of quantum paradoxes can be observed in joint measurements when the resolution is sufficiently reduced~\cite{Virzi}. Given this caveat, it should not have come as a surprise that post-selected weak measurements can result in values outside of the spectrum of eigenvalues \cite{AAV88}. By defining the precise interference between all of the eigenstates contributing to the post-selection probability, post-selection ensures that a strong measurement of the target observable is incompatible with the observation of its weak values. The weak value itself is determined by the quantum interference effects that occur in this specific combination of initial state and post-selected final outcome. The weak value captures an important aspect of the relationship between the measurement used in the post-selection and the physical property observed in the weak measurement. In doing so, it provides a possible resolution to quantum paradoxes~\cite{Aha+02,LS09,Yok+09,Hance2023QCC} and should therefore be considered a reasonable candidate for the values of a physical property conditioned by the strong measurement of a different observable. However, there is a somewhat understandable discomfort associated with the idea that a physical property should have well-defined values that cannot be observed in an isolated single shot measurement, resulting an an ongoing controversy regarding the proper interpretation of weak values~\cite{Vai17,Mat19,Hance2023WeakValues,Vaidman2023CommentHance,Hance2023ReplyVaidman}. This controversy has its roots in the objection to the use of post-selection in a situation where individual outcomes cannot be resolved, making it difficult to distinguish between statistical correlations and physical effects~\cite{Leg89,Per89}. It has even been argued that anomalous weak values might arise from a direct post-selection of extremal meter outcomes, with no relation to the physics of the system~\cite{FC14}. This rather provocative claim has been refuted by a number of arguments rejecting the underlying assumption that our knowledge of the physics of measurements is limited to the statistics observed in the output of what is erroneously treated as a black box~\cite{AR14,Coh14,HIS14,Brodutch2015CommentFerrie,Dre15,Sok15,Rom+16,MO16}. 

As part of this debate, it has been pointed out that the weak value describes a quantum interference effect that happens within the system and is therefore independent of the meter statistics~\cite{Dre15}. This observation is consistent with the identification of weak values as conditional averages of the Kirkwood-Dirac distribution of the input state~\cite{Kir33,Dir45,Arv+24}, where anomalous weak values are linked to the negativity of the quasi-probability distribution. Quantum interference effects can thus be represented by negative quasi-probabilities in the distribution of eigenvalues, suggesting that the weak value corresponds to a conditional expectation value rather than to the value for a single measurement~\cite{Wis02,DAJ10,DJ12,Dre+14,Ips15}. However, it is not at all obvious what the physical meaning of quasi-probabilities is. The theory of weak measurements suggests that, for pure states, the meter shifts have a well-defined value, corresponding more closely to the effects of an eigenvalue than to the effects of an expectation value~\cite{AB05,Vai+17}. It may therefore be necessary to take a closer look at the statistics of weak measurements to decide whether weak values represent the physics of individual systems or not.

A noteworthy result, somewhat overlooked in the discussion so far, is the surprising discovery that the mathematical description of measurement uncertainties introduced by Ozawa in 2003 identifies the weak value of an observable as an error-free estimate of the observable, obtained from the combination of prior information given by a pure state input and the information obtained from the final measurement result~\cite{Oza03,Hal04}. This implies that any precise measurement performed on a pure state resolves all uncertainties of the initial state in terms of the weak values defined by the initial state and the measurement outcome. This interpretation of weak values as contextualised quantum fluctuations of the input state is consistent with the mathematical structure of the Hilbert space formalism. As shown by Hall \textit{et al}~\cite{Hall16}, the variance of the weak values obtained for the different measurement outcomes of any fully resolved measurement is exactly equal to the uncertainty of the observable in the initial state, leaving no room for any additional fluctuations in the post-selected weak values. It can be conjectured that the force acting on the meter system in a weak measurement should not fluctuate. In 2021, one of the authors showed that it is possible to measure the fluctuations of the force on the meter by compensating its effects using an appropriate unitary operation \cite{Hof21}. It is then possible to confirm experimentally that the Ozawa-Hall uncertainties accurately describe the fluctuation of forces acting on the meter in a weak measurement \cite{Lem+22,Dvo+25,Fukuda}. This provides compelling evidence that weak values accurately describe the distance by which the measurement interaction shifts a quantum meter in a weak measurement statistically conditioned by a specific combination of initial and final conditions~\cite{MH23}. The interpretational confusion mainly arises from the difficulty of separating fluctuations originating from the initial meter state from the intrinsic fluctuations of the observable conditioned by the post-selected outcome. An initial investigation of meter fluctuations in weak measurements showed that post-selection can actually reduce the fluctuations of the initial meter state \cite{Par11,Par14,Ber15}, corresponding to a negative weak variance of the observable~\cite{Ogawa}. However, two of the authors later showed that this effect originates from quantum correlations between the meter fluctuations and the post-selected measurement outcome~\cite{MH24}. Since the probability of a successful post-selection depends on the magnitude of the measurement back action, the meter statistics are updated, causing meter-state-dependent changes to the variance of the meter readout.

In this paper, we investigate this relation between updates of the meter statistics due to the information gained in the post-selection process and the quantum interferences between different eigenstates of the target observable defined by the post-selection of a final state of the system. We identify the precise mapping between the interferences between eigenstates of the system and the interferences between meter wavefunctions shifted by the corresponding eigenvalues of the system. This mapping is expressed by a specific quasi-probability of the post-selected system that describes the average meter shift associated with interferences between two eigenstates by the average of the two eigenvalues associated with these states. The associated meter statistics is given by an interference between two meter wavefunctions shifted by an amount proportional to the eigenvalues of the target observable. These interference terms describe the measurement strength dependence of post-selected measurements. It is therefore important to understand how the contribution of these interference terms to the post-selected meter statistics changes with increasing measurement strength. We analyse the measurement strength dependence by identifying the linear and quadratic dependence in the weak measurement limit. The quadratic term is essential, since it should describe the effects of quantum uncertainties of the target observable on the statistics of the meter readout. The goal of this work is to develop physically motivated criteria that allow us to distinguish between statistical updates of the meter caused by the post-selection and dynamical effects that transfer variances of the target observable to the meter by linear shifts of the readout observable. Our results show that it is sufficient to require the positivity of the objective conditional fluctuations of the physical property in the system. Under this constraint, the conditional fluctuations of the system observable are given by Ozawa-Hall uncertainties \cite{Oza03,Hal04}, consistent with our previous results on quantum fluctuations in weak measurement interactions \cite{Hof21,MH24}. Remarkably, this positive conditional uncertainty is obtained by attributing a negative statistical weight to the diffusive change of the meter statistics associated with quantum interferences. This result highlights the fact that quantum interferences cannot be interpreted as separate realities even when their quasi-probabilities are positive. Perhaps the most significant result of the analysis is the term that describes a Bayesian update of the meter statistics associated with the dephasing effect induced by the meter back action. The origin of this Bayesian update is the dependence of the post-selection probability on the squared meter momentum. In the weak measurement regime, the position dependent value of squared meter momentum is described by the corresponding weak value.

In summary, our results show that the quantum formalism explains the changes to the meter statistics in the weak measurement regime in terms of a specific combination of meter diffusion caused by intrinsic conditional uncertainties of the target observable and a Bayesian update of the statistics associated with the change of post-selection probability by the back action of the meter on the system. The reason why this relation is not immediately obvious is the role that quantum interference plays in the modification of post-selection effects. As measurement strength increases, dephasing changes the meaning of post-selection in a fundamental way, limiting the present analysis to the weak measurement regime. Nevertheless the insights presented here are essential for a fundamental understanding of quantum measurements, highlighting the reasons why the statistical contributions of interference effects should not be attributed to hidden realities. We understand that it might seem somewhat disturbing that it is possible to identify diffusion terms and Bayesian updates without a well-defined Kolmogorovian set of events. However, our results demonstrate that the operator algebra replaces the classical framework with a consistent alternative ensuring the positivity of observable statistics by other means. Ultimately, quantum mechanics can only be understood in terms of the dynamics described by the interactions needed to obtain observable effects.

This paper is organised as follows. In \Cref{sec:model}, we analyse the meter readout statistics of a post-selected measurement, and establish the relation between the quantum interference terms that appear in the post-selection probability, and the quantum interference terms between the corresponding shifts of the meter wavefunction. The different normalised contributions to the post-selected meter state are expressed by a quasi-probability distribution, where the meter shift of the interference terms is given by the average of the two eigenvalues involved in the interference. In \Cref{sec:fluctuations}, we consider the relation between the variance of this quasi-probability and the variance of the conditional readout distribution. We show that the contribution of the interference terms depends on measurement strength, indicating a modification of the readout statistics dependent on the meter state. Based on previous research, we identify a contribution to the meter readout fluctuations that originates from the back action dynamics of the system-meter interaction. In \Cref{sec:analysis}, we examine the precise ways the shape of the meter interference pattern changes as measurement strength increases.  We identify a Bayesian update that depends on the meter state, related to the weak value of squared meter momentum at each readout position of the meter. This update takes into account that the post-selection probability depends on the value of squared meter momentum. Effectively, the post-selection modifies the meter statistics based on the correlations between the post-selected outcome and the initial meter positions before the measurement interaction was applied. This modification is not related to the statistics of the observable in the system. However, there is an additional modification of the interference patterns represented by a negative diffusion term. This term represents the modification of the statistics of quantum interference from the non-positive quasi-probability introduced in \Cref{sec:model} to a positive conditional fluctuation represented by Ozawa-Hall uncertainties. In \Cref{sec:nonBayesian}, we point out that the reason why the interference patterns of Gaussian meter states do not change as measurement strength increases is the apparent cancellation between a Bayesian update (that broadens the distribution) and the negative diffusion (which narrows it). To look outside this cancellation regime, we consider a meter wavefunction defined by a cosine. For this wavefunction, there is no Bayesian update, since the weak value of squared meter momentum is independent of readout position. With this meter wavefunction, the Ozawa-Hall uncertainty of the observable appears as a noise background that reduces the contrast of the meter distribution. In \Cref{sec:meterinterference}, we summarise our findings, and discuss their implications for our understanding of the measurement process. \Cref{sec:conclusions} concludes the paper.

\section{Quantum coherence of the meter system in a post-selected measurement}
\label{sec:model}

Let us consider the measurement of a system observable $\hat{A}$. According to the established formalism, the observable is represented by a self-adjoint operator with eigenstates $\ket{a}$ and the associated eigenvalues $A_a$. The following analysis will always refer to a complete orthogonal basis $\{\ket{a}\}$, but we would like to note that there is no requirement that the eigenvalues of different $a$ must be different. Possible degeneracies can be accounted for by letting $A_{a}-A_{a^\prime}$ go to zero. The interaction that transfers quantitative information about $\hat{A}$ from the system to the meter should be described in terms of a unitary operation acting on the joint quantum state of the system and the meter. In the following, we consider the standard von Neumann model of a continuous variable meter, where the measurement interaction results in a displacement of the meter’s initial wavefunction $\phi(x)$ by an amount proportional to the value of $\hat{A}$. This standard form of the system-meter interaction can be expressed using the momentum operator $\hat{p}$ as the generator of the displacement,
\begin{equation}
\label{eq:interaction}
\hat{U}=\exp\left(-\frac{i}{\hbar}s\hat{A}\otimes\hat{p}\right),
\end{equation}
where $s$ denotes the effective strength of the measurement interaction. The effect of this conditional displacement on the initial meter state $\ket{\phi}$ can be written as
\begin{equation}
\bra{x}\hat{U}\ket{\phi}=\sum_a\phi(x-sA_a)\ketbra{a}{a},
\end{equation}
where the quantum mechanics of the system is represented by the projection operators onto the eigenstates $\ket{a}$ associated with the eigenvalues $A_a$ of $\hat{A}$.

The choice of a continuous variable meter is an essential part of our analysis. On the one hand, the readout results of measurement position $x$ describe macroscopically observable outcomes, on the other hand the coherent wavefunctions describe the full system-meter entanglement in terms of conditional quantum interference effects. If the system is in a coherent superposition of different eigenstates $\ket{a}$ of $\hat{A}$, the interaction generates an entangled state, where each eigenstate component $\ket{a}$ is correlated with a meter state shifted by $s A_a$. When the system is omitted, the meter state appears to be a statistical mixture of different meter shifts $s A_a$, suggesting a classical interpretation of the meter dynamics where each eigenvalue $A_a$ represents a separate value of $\hat{A}$. However, this interpretation fails if a subsequent measurement is performed on the system and the result $\bra{f}$ of this measurement is itself represented by a superposition of different eigenstates $\ket{a}$. The post-selection of a specific measurement outcome $\bra{f}$ defines quantum coherences between different meter displacements. These coherences are associated with quantum interference between the corresponding eigenstates of $\hat{A}$ that contribute to the post-selection probability $P(f)$. The joint probability of post-selection and a meter readout of $x$ is given by
\begin{equation}
\label{eq:joint probability}
P(f,x)=\sum_{a,a'}\braket{f|a}\hskip-1mm\bra{a}\hat{\rho}\ket{a'}\hskip-1mm\braket{a'|f}\phi(x-sA_a)\phi^*(x-sA_{a'}),
\end{equation}
where quantum interference between the eigenstates of $\hat{A}$ in the initial system state $\hat{\rho}$ necessarily implies quantum interference between the corresponding displaced wavefunctions of the meter. The post-selection probability $P(f)$ is obtained by marginalising \Cref{eq:joint probability} over the distribution of meter readouts $\{x\}$,
\begin{equation}
\label{eq:marginal probability}
P(f)=\sum_{a,a'}\braket{f|a}\hskip-1mm\bra{a}\hat{\rho}\ket{a'}\hskip-1mm\braket{a'|f} \left(1-R(a,a')\right),
\end{equation}
where
\begin{equation}
R(a,a')=1-\int\phi(x-sA_a)\phi^*(x-sA_{a'})\mathrm{d}x
\end{equation}
describes the decoherence between eigenstates of $\hat{A}$ caused by the measurement interaction. The terminology used here emphasises that one minus the overlap of differently displaced meter wavefunctions represents the distinguishability of the meter wavefunctions, and therefore doubles as a measure of resolution for the eigenstates $\ket{a}$ and $\ket{a^\prime}$~\cite{Kartik}. \Cref{eq:joint probability} therefore describes the tight correlations between quantum interference in the system eigenstates $\{\ket{a}\}$ and quantum interference in the meter states that represent the corresponding displacements $s A_a$. 

Meter interference is a phenomenon that appears only in post-selected measurement statistics. The conditional probability density for the meter readout of a post-selected final outcome $f$ is given by
\begin{equation}
\label{eq:conditional probability}
P(x|f)=\sum_{a,a'}\frac{\braket{f|a}\hskip-1mm\bra{a}\hat{\rho}\ket{a'}\hskip-1mm\braket{a'|f}}{P(f)}\phi(x-sA_a)\phi^*(x-sA_{a'}).
\end{equation}
The formalism clearly separates meter shifts associated with eigenvalues from statistical contributions of interferences between differently displaced meter wavefunctions. Although this is an obvious consequence of the standard quantum formalism, it is not usually recognized that this appearance of quantum interference on the meter side of the measurement represents a fundamental change of the physics of measurement. The purpose of a measurement is to reveal the value of the observable $\hat{A}$. As long as each meter readout can be traced back to a specific eigenvalue, we can safely accept that all correct measurement outcomes can be represented by eigenvalues. However, quantum mechanics disrupts this convenient assumption by associating part of the meter statistics with interference effects involving two different eigenvalues. This is the fundamental reason why the outcomes of weak measurements cannot be explained by eigenvalues. In the following, we will show how the quantum statistics of meter interference relates to quasi-probabilities, post-selected quantum uncertainties, and quantum correlations of the initial meter state.

We start our analysis by considering the contribution of meter interference to the conditional mean value of the position $\hat{x}$,
\begin{equation}
\label{eq:mean value of x}
\braket{\hat{x}}_f=\sum_{a,a'}\frac{\braket{f|a}\hskip-1mm\bra{a}\hat{\rho}\ket{a'}\hskip-1mm\braket{a'|f}}{P(f)}\int x\;\phi(x-sA_a)\phi^*(x-sA_{a'})\mathrm{d}x.
\end{equation}
The meter wavefunction $\phi(x)$ determines the distribution of measurement errors. Since asymmetric errors will result in a systematic bias, it is reasonable to assume that the meter wavefunction will be symmetric around a centre we can define at $x=0$, meaning positive and negative deviations from this central position should be equally likely,
\begin{equation}
\label{eq:symmetric meter state}
\phi(x)=\phi^*(-x).
\end{equation}
The interference pattern will then be symmetric around the average shift of the two contributions, so that the contribution of the interference terms between different meter displacements to the average meter readout can be expressed in terms of the average of the two interfering eigenvalues,
\begin{equation}
\label{eq:interference_average}
\int x\;\phi(x-sA_a)\phi^*(x-sA_{a'})\mathrm{d} x= s\frac{A_a+A_{a'}}{2}(1-R(a,a')).
\end{equation}
The conditional mean value of the meter readout $\hat{x}$ then reduces to
\begin{equation}
\braket{\hat{x}}_f=s\sum_{a,a'}\frac{A_a+A_{a'}}{2}Q(a,a'|f),
\end{equation}
where the average meter shifts are described by individual contributions given by averages of eigenvalues, with a quasi-probability of
\begin{equation}
Q(a,a'|f)= \frac{\mbox{Re}\left(\braket{f|a}\hskip-1mm\bra{a}\hat{\rho}\ket{a'}\hskip-1mm\braket{a'|f}\right)}{P(f)}(1-R(a,a')).
\end{equation}
This quasi-probability represents a distribution over pairs of eigenvalues $(A_a,A_{a'})$ and obeys the normalisation condition
\begin{equation}
\sum_{a,a'}Q(a,a'|f)=1.
\end{equation}
 This quasi-probability describes the mean conditional meter readout in terms of the average eigenvalues associated with each element of the density matrix. However, $Q(a,a'|f)$ takes negative values whenever the contribution of the interference between $\ket{a}$ and $\ket{a^\prime}$ to the post-selection probability $P(f)$ is negative. 

Even though the quasi-probability $Q(a,a'|f)$ is derived from the system-meter interaction, it does not depend on any specific properties of the meter. The entire measurement strength dependence is contained in the resolution $R(a,a')$, and this term relates to the loss of coherence caused by the measurement back action. In the weak limit ($s \to 0$), the resultion is negligible ($R(a,a')\to 0)$) and the quasi-probability describes the intrinsic statistics of the initial state $\hat{\rho}$. This is the first major result of our analysis - the quasi-probability $Q(a,a'|f)$ is an objective description of the causes of meter interference within the post-selected quantum system. Note that "objective" does not mean "objectively real", but refers to the intrinsic quantum statistics of the object. This distinction becomes relevant because the system-meter entanglement cannot be described by objective statements, with or without realist assumptions. Even though the combinations of eigenvalues associated with $(a,a')$ do not represent any physical events or outcomes, the quasi-probability $Q(a,a'|f)$ provides a reliable description of the contributions of quantum interference to the observable statistics of meter readouts.

The average meter readout provides an objective value of the system observable $\hat{A}$ conditioned by the post-selection of $f$ that is given by the average of the quasi-probability $Q(a,a'|f)$,
\begin{equation}
\label{eq:meanAf}
\braket{\hat{A}}_f=\sum_{a,a'}\frac{A_a+A_{a'}}{2}Q(a,a'|f).
\end{equation}
In this formula, quantum interference terms are treated as independent statistical contributions, where the value of $\hat{A}$ is given by the average of the two eigenvalues involved in the interference. For sufficiently small measurement interaction strength $s$, the resolution $R(a,a')$ approaches zero and the contributions of coherences between different eigenstates in $Q(a,a'|f)$ are undiminished. The assignment of average values $(A_a+A_{a'})/2$ to off-diagonal elements of the density matrix then results in a conditional value of $\hat{A}$ given by
\begin{equation}
\label{eq:weak limit of mean value}
\left.\braket{\hat{A}}_f\right|_{s=0}=\textrm{Re}\left(\frac{\bra{f}\hat{A}\hat{\rho}\ket{f}}{\bra{f}\hat{\rho}\ket{f}}\right).
\end{equation}
As expected, this is the real part of the weak value, experimentally observed in post-selected weak measurements characterised by sufficiently low measurement strength $s$ \cite{AAV88}. Importantly, within the present framework, the real part of the weak value emerges naturally from the non-classical statistics of the quasi-probability $Q(a,a'|f)$, where interference terms are considered as independent measurement outcomes with a meter shift given by the average of the two eigenvalues of $\hat{A}$. 

In this first result of our analysis, we have shown that, in the weak measurement limit ($s \to 0$), the meter interference effects of post-selected measurements can be mapped onto an objective quasi-probability $Q(a,a'|f)$ that associates interferences between different eigenstates in the system with the average of their two eigenvalues. In this sense, weak values can be considered as quantum interference phenomena~\cite{Dre15}. However, it remains unclear how one should understand the identification of destructive interference with the appearance of negative quasi-probabilities. Indeed, \Cref{eq:conditional probability} suggests that interference terms should not be interpreted as meter statistics shifted by a single value of $(A_a+A_{a'})/2$ since the probability distribution of each interference term is quite different from the distributions of the diagonal elements of the density matrix. In order to understand the statistics of weak measurements properly, we need to examine the way in which the contribution of interference terms to the meter statistics changes as measurement strength increases. 

\section{Fluctuations of the meter readout}
\label{sec:fluctuations}

The conditional average of the meter statistics provides an estimate of the value of $\hat{A}$ based on the expectation that the change of the meter statistics is caused by meter shifts proportional to the value of $\hat{A}$. Likewise, one can expect that the statistical fluctuations of the meter statistics provide an estimate of the conditional fluctuations of $\hat{A}$ \cite{Ogawa,MH24}. Without post-selection, the uncertainty $\Delta A^2$ would cause a proportional increase in the meter variance, corresponding to a diffusive broadening of the meter statistics,
\begin{equation}
\label{eq:basic}
    \Delta x^2 = s^2 \Delta A^2 + \int x^2 |\phi(x)|^2 dx. 
\end{equation}
Note that the fluctuations of the initial meter statistics are not changed by the shifts in the value of $x$ caused by the system-meter interaction. In the absence of post-selection, the quadratic dependence of meter uncertainty on measurement strength clearly distinguishes the objective contribution of the system uncertainty $\Delta A^2$ from the uncertainty contributed by the initial meter state. 

We can now apply the same analysis to the conditional meter fluctuations given by $P(x|f)$ in Eq.(\ref{eq:conditional probability}) above. The quasi-probability $Q(a,a'|f)$ can be used to define a variance of the meter shifts given by
\begin{equation}
\label{eq:VarA}
V_Q=\langle\hat{A}^2\rangle_f - \langle\hat{A}\rangle_f^2 = \sum_{a,a'}\left(\frac{A_a+A_{a'}}{2}\right)^2Q(a,a'|f)-\left(\sum_{a,a'}\frac{A_a+A_{a'}}{2}Q(a,a'|f)\right)^2
\end{equation}
In the conditional meter statistics, this variance replaces the unconditional uncertainty $\Delta A^2$. However, it should be kept in mind that $V_Q$ is the variance of the non-positive quasi-probability $Q(a,a'|f)$ and may take on negative values itself. 

In addition to the variance $V_Q$, it is necessary to determine the variances of each interference term. These can be described by normalised distributions centred around $x=0$,
\begin{equation}
I(a,a',x)=\frac{\phi\left(x-s\frac{A_a-A_{a'}}{2}\right)\phi^*\left(x+s\frac{A_a-A_{a'}}{2}\right)}{1-R(a,a')}.
\end{equation}
For $a=a'$, this distribution is given by the initial meter statistics, $I(a,a,x)=|\phi(x)|^2$. For $a \neq a'$, the interference pattern changes with increasing measurement strength. The fluctuations of the meter readout are then given by
\begin{equation}
\label{eq:fluctuation}
\Delta x^2_f= s^2 V_Q + \sum_{a,a'}Q(a,a'|f)\int x^2 I(a,a',x)\mathrm{d}x.
\end{equation}
In the weak measurement limit, the interference patterns $I(a,a',x)$ are very similar to the original meter statistics $|\phi(x)|^2$, with only small changes that are quadratic in $s$. Based on the observation that the unconditional uncertainty $\Delta A^2$ determines the quadratic increase of readout fluctuations with measurement strength, it is possible to define the weak conditional uncertainty of $\hat{A}$ as
\begin{equation}
    \Delta A^2_f = \frac{1}{2} \left. \frac{\partial^2}{\partial s^2} \Delta x_f^2 \right|_{s=0}.
\end{equation}
This definition corresponds to the one used by Ogawa et al. \cite{Ogawa} to identify negative weak uncertainties for Gaussian meter states. Our analysis shows that this result depends on the specific meter state,
\begin{equation}
    \Delta A^2_f = V_Q + \sum_{a,a'} Q(a,a'|f) \left. \frac{1}{2}\frac{\partial^2}{\partial s^2} \int x^2 I(a,a',x) dx \right|_{s=0}.  
\end{equation}
The integral can be solved for all $(a,a')$ by
\begin{equation}
\label{eq:Kfactor}
\frac{1}{2}\left. \frac{\partial^2}{\partial s^2} \int x^2 I(a,a',x) dx \right|_{s=0} = \left(\frac{A_a-A_{a'}}{2}\right)^2 \; K_Q.
\end{equation}
where $K_Q$ is a characteristic value of the meter wavefunction, independent of the system property $\hat{A}$. This is the second significant result of our analysis - the contribution of meter interference to the readout fluctuations of the measurement has a quadratic measurement strength dependence that can be factorized into a universal dependence on the squared difference between the eigenvalues and a value $K_Q$ that characterises the initial state of the meter and is completely independent of the physics of the system. For Gaussian meter states, $K_Q=0$ and the weak conditional uncertainty is equal to the variance $V_Q$. This variance describes the change of meter noise in post-selected measurements reported in previous work \cite{Par11,Par14,Ber15} and was first interpreted as a weak uncertainty in \cite{Ogawa}. Later, two of us showed that this result depends on the correlation between position and momentum in the meter state and includes a statistical update of the meter state by the post-selection of $f$ \cite{MH24}. Negative values of weak conditional uncertainties only appear as a result of such correlations between position and momentum in the meter state and should not be attributed to the objective conditional uncertainties of the observable $\hat{A}$ in the system.

The ambiguity of the weak conditional uncertainty $\Delta A_f^2$ introduced by the meter state characteristic $K_Q$ indicates that the quasi-probability variance $V_Q$ is only one of a whole class of quantum conditional uncertainties defined by different operator orderings. In terms of the quasi-probabilities $Q(a,a'|f)$, the conditional weak fluctuations of $\hat{A}$ are given by
\begin{eqnarray}
\label{eq:KQ}
    \Delta A_f^2 &=& \left. V_Q \right|_{s=0} + K_Q \; \sum_{a,a'} \left(\frac{A_a-A_{a'}}{2}\right)^2 Q(a,a'|f)
\nonumber \\
    &=& \sum_{a,a'} \left(\frac{(1+K_Q)(A_a^2+A_{a'}^2)+2 (1-K_Q) A_a A_{a'}}{4}\right) Q(a,a'|f) - \left. \langle \hat{A} \rangle_f^2 \right|_{s=0}
\end{eqnarray}
As noted above, the quasi-probability variance $V_Q$ is obtained for $K_Q=0$, which happens to be the characteristic value of Gaussian meter states. The variance $V_Q$ is determined by assigning the eigenvalue average $(A_a+A_{a'})/2$ to each density matrix element in the calculation of the average square, where the quasi-probability $Q(a,a'|f)$ represents the statistical weight of each density matrix element. $K_Q$ modifies this contribution of the interference terms to the average of the squared values of $\hat{A}$. Effectively, $K_Q$ assigns fluctuations to the average value of $(A_a+A_{a'})/2$ of each interference term. For example, a value of $K_Q=1$ removes the contributions of the products $A_a A_{a'}$ in \Cref{eq:KQ}, leaving only a sum over the squares of the eigenvalues. This sum corresponds to the assumption that each interference term represents a statistical distribution of two eigenvalues with a probability of $1/2$ for each. In general, these modified fluctuations indicate that it is not possible to assign specific values of $\hat{A}$ to the elements of the quasi-probability $Q(a,a'|f)$. It may therefore be useful to express \Cref{eq:KQ} as an operator relation,
\begin{equation}
\label{eq:qvargeneral}
    \Delta A_f^2 
=  \left(\frac{1+K_Q}{2} \mbox{Re} \left(\frac{\bra{f}\hat{A}^2\hat{\rho}\ket{f}}{\bra{f}\hat{\rho}\ket{f}}\right) + \frac{1-K_Q}{2} 
\frac{\bra{f}\hat{A}\hat{\rho}\hat{A}\ket{f}}{\bra{f}\hat{\rho}\ket{f}}\right)
-\left(\textrm{Re}\left(\frac{\bra{f}\hat{A}\hat{\rho}\ket{f}}{\bra{f}\hat{\rho}\ket{f}}\right)\right)^2.
\end{equation}
This lets us identify the ambiguity of the conditional weak uncertainty with an operator ordering problem. 

As explained above, there is a clear physical explanation of the ambiguity. As suggested in \cite{FC14}, post-selection can update the meter statistics in a manner unrelated to the actual meter shift. This kind of Bayesian update of the meter fully explains all negative values of conditional weak uncertainties \cite{MH24}. The meter wavefunction determines the amount and the sign of the Bayesian contribution through $K_Q$. 

If the Bayesian contribution vanishes, all negative conditional weak uncertainties also vanish. The remaining conditional uncertainty represents the actual fluctuations in the value of the physical property $\hat{A}$ that determines the meter shifts. Since the weak values of $\hat{A}^2$ are unbounded, possibly taking any value between negative infinity and infinity, negative conditional weak uncertainties only vanish for $K_Q=-1$ and the conditional uncertainties of the meter shifts are given by
\begin{equation}
\label{eq:ozawahall}
\varepsilon^2_A(f)=\frac{\bra{f}\hat{A}\hat{\rho}\hat{A}\ket{f}}{\bra{f}\hat{\rho}\ket{f}}-\left(\textrm{Re}\left(\frac{\bra{f}\hat{A}\hat{\rho}\ket{f}}{\bra{f}\hat{\rho}\ket{f}}\right)\right)^2.
\end{equation}
This definition of conditional uncertainties was motivated by the measurement theory of Ozawa \cite{Oza03}, which was interpreted as an estimation problem by Hall in \cite{Hal04}, where it appears as an estimation error for estimates of $\hat{A}$ based on the measurement outcomes $f$. It has also been shown that Ozawa-Hall uncertainties can be observed as physical fluctuations of the parameter in the unitary transformation that determines the rotation of a qubit probe \cite{Hof21,Lem+22,Dvo+25,Fukuda}. According to the result in \cite{MH24}, $K_Q$ evaluates correlations between position and momentum. At $K_Q=-1$, the information on meter momentum obtained in the post-selection has no effect on the meter fluctuation.

The difference between the conditional weak uncertainty observed with a meter state characterised by $K_Q$ and the Ozawa-Hall uncertainty is given by
\begin{equation}
\label{eq:qvar}
    \Delta A_f^2 - \varepsilon^2_A(f)
=  \frac{1+K_Q}{4} \left(\frac{\bra{f} (\hat{A} (\hat{A}\hat{\rho}-\hat{\rho}\hat{A})- (\hat{A}\hat{\rho}-\hat{\rho}\hat{A})) \hat{A}\ket{f}}{\bra{f}\hat{\rho}\ket{f}}\right).
\end{equation}
As pointed out in \cite{MH24}, the double commutation relation that defines the difference between the conditional meter statistics and the corresponding Ozawa-Hall uncertainty describes the sensitivity of the post-selection probability to the back action of a weak measurement. In the measurement interaction, the meter momentum $\hat{p}$ conditions a change of the system state, resulting in statistical correlations between meter momentum and the post-selected outcome. This correlation means that the post-selection of an outcome $f$ modifies the quantum statistics of the initial meter state by providing information about the magnitude of $\hat{p}^2$. In the following, we will show that this Bayesian update of the meter statistics is described by the measurement strength dependent change of the interference patterns.

\section{Analysis of meter interference}
\label{sec:analysis}

We will now get to the main result of this work. As the preliminary analysis showed, the interference patterns $I(a,a',x)$ combine an effect of the objective quantum statistics of $\hat{A}$ with a meter state dependent Bayesian update of the outcome distribution $x$. The Bayesian update is related to the effects of meter momentum $\hat{p}$ on the post-selection probability $P(f)$, and the objective effects of $\hat{A}$ describe an uncertainty in the meter shifts. The latter can only take the form of a diffusive change of the meter distribution, since random meter shifts resemble a diffusion process in the value of meter position.

For $s=0$, the normalised distribution of the symmetric meter interference $I(a,a',x)$ is equal to the initial distribution of the meter readout, $P(x)=|\braket{x|\phi}|^2$. As the measurement interaction strength $s$ increases, the normalised distributions representing quantum interferences $(a \neq a')$ change their shape. Because $I(a,a',x)$ is symmetric about $x=0$, the lowest order contribution to this change is of second order in the measurement interaction strength $s$. In the weak limit, the effects of quantum interference on the readout statistics is given by
\begin{equation}
\label{eq:2deriv}
\frac{1}{2}\left.\frac{\partial^2}{\partial s^2} I(a,a',x)\right|_{s=0}=-\frac{1}{\hbar^2}\left(\frac{A_a-A_{a'}}{2}\right)^2 \left(\textrm{Re}\left(\frac{\bra{x}\hat{p}^2\ket{\phi}}{\braket{x|\phi}}\right)+\left|\frac{\bra{x}\hat{p}\ket{\phi}}{\braket{x|\phi}}\right|^2
-2\braket{\hat{p}^2}\right) P(x).
\end{equation}
Here, all derivatives in $x$ have been expressed in terms of the momentum operator $\hat{p}$, where
\begin{equation}
    \bra{x} \hat{p} \ket{\phi} = -i \hbar \frac{\partial}{\partial x} \braket{x|\phi}.
\end{equation}
In addition, we assume that the meter wavefunction has a momentum expectation value of zero ($\braket{\hat{p}}=0$). 
Quantum mechanics then lets us relate the change of the interference pattern to the quantum statistics of momentum given by the differences between weak values of $\hat{p}$ and the expectation value of $\hat{p}^2$ in the initial meter state $\ket{\phi}$. 

Let us start with the effects of $\hat{p}^2$ on the post-selection probability $P(f)$. In general, the amount of decoherence induced by the back action is proportional to $\hat{p}^2$, so the contribution by a readout of $x$ to the interference pattern will be decreased if the local value of $\hat{p}^2$ at $x$ is larger than $\braket{\hat{p}^2}$, and increased if it is smaller than $\braket{\hat{p}^2}$. 
Post-selection effectively performs a weak measurement of $\hat{p}^2$, modifying the probability of the outcome by
\begin{equation}
   \left. \frac{1}{2} \frac{\partial^2}{\partial s^2} P(f|x) \right|_{\mathrm{BA}}= \mbox{Re}\left(\frac{\bra{x}\hat{p}^2\ket{\phi}}{\braket{x|\phi}}\right) \frac{1}{2} \frac{\partial^2}{\partial \phi_A^2} P(f) ,
\end{equation}
where $BA$ indicates that we are only considering the dynamical change of $P(f)$, not the Bayesian update of $P(f)$ associated with measurement information relating to $\hat{A}$. The sensitivity of the post-selection probability $P(f)$ to dynamics generated by $\hat{A}$ is given by
\begin{equation}
\label{eq:sense}
\frac{\partial^2}{\partial \phi_A^2} P(f)  = -\frac{1}{\hbar^2}\bra{f} \left(\hat{A} (\hat{A}\hat{\rho}-\hat{\rho}\hat{A})- (\hat{A}\hat{\rho}-\hat{\rho}\hat{A})\hat{A}\right)\ket{f}.
\end{equation}
The contributions of each interference term $I(a,a',x)$ to $P(f)$ is given in \Cref{eq:marginal probability}. The measurement interaction modifies $P(f)$ by reducing the coherence between the eigenstates according to the increase in resolution $R(a,a')$ given by
\begin{equation}
    \frac{1}{2} \left. \frac{\partial^2}{\partial s^2} R(a,a') \right|_{s=0}= \frac{2}{\hbar^2} \braket{\hat{p}^2} \left(\frac{A_a-A_{a'}}{2}\right)^2.
\end{equation}
This equation describes the measurement strength dependence of the normalization of the interference pattern $I(a,a',x)$. The contribution of the Bayesian update to the right hand side of Eq.(\ref{eq:2deriv}) is then given by
\begin{equation}
\label{eq:Bayes}
B(x) = -\frac{2}{\hbar^2} \left(\textrm{Re}\left(\frac{\bra{x}\hat{p}^2\ket{\phi}}{\braket{x|\phi}}\right)-\braket{\hat{p}^2}\right) P(x).
\end{equation}
This result shows in detail how quantum interference between different meter shifts describes the Bayesian update of the meter statistics due to momentum information gained in the post-selection.

In addition to this Bayesian update, meter interference describes an objective effect of the fluctuations in the physical property $\hat{A}$. Since these fluctuations introduce random meter shifts, the change of the meter distribution is given by a diffusion term,
\begin{equation}
\label{eq:diffusion}
\frac{1}{2} \frac{\partial^2}{\partial x^2} P(x) = - \frac{1}{\hbar^2}\left(
\textrm{Re}\left(\frac{\bra{x}\hat{p}^2\ket{\phi}}{\braket{x|\phi}}\right)
- \left|\frac{\bra{x}\hat{p}\ket{\phi}}{\braket{x|\phi}}\right|^2\right) P(x).
\end{equation}
This diffusion term describes the effect of a random shift of the meter distribution, where the operator $\hat{p}$ serves as the generator of the spatial displacements. We can now decompose the change of the interference pattern into a Bayesian contribution and a diffusion term that modifies the intrinsic statistics of $\hat{A}$,
\begin{equation}
\label{eq:decomposition}
\frac{1}{2}\left.\frac{\partial^2}{\partial s^2}I(a,a',x)\right|_{s=0}=  \left(B(x) - \frac{1}{2} \frac{\partial^2}{\partial x^2} P(x) \right) \left(\frac{A_a-A_{a'}}{2}\right)^2.
\end{equation}
This is the main result of our analysis. The contribution of each quantum interference term to the meter statistics contributes both to a Bayesian update of the meter statistics due to momentum information obtained in the post-selection, and to the diffusion effect caused by the objective randomness of the observable $\hat{A}$ in the system. The negative sign in front of the diffusion term indicates that the assignment of averages of the eigenvalues to interference terms overestimates the fluctuations when $Q(a,a'|f)$ is positive, and underestimates them when it is negative. In the latter case, the negative sign is multiplied by a negative quasi-probability, resulting in a positive correction that ensures a positive diffusion effect in the experimentally observable meter distribution.

The contribution of the Bayesian update to the readout fluctuations is given by
\begin{equation}
    \int x^2 B(x) dx = -\frac{2}{\hbar^2}\left(\mbox{Re}\left(\bra{\phi}\hat{x}^2\hat{p}^2\ket{\phi}\right) - \braket{\hat{x}^2}\braket{\hat{p}^2}\right) 
\end{equation}
Additionally, the negative diffusion reduces the fluctuations contributed by quantum interference by a constant value of
\begin{equation}
   - \int x^2 \frac{1}{2}\frac{\partial^2}{\partial x^2} P(x) = -1.
\end{equation}
Since the negative diffusion effect is independent of the initial meter state, the value of $K_Q$ can be expressed entirely in terms of the correlation between $\hat{x}^2$ and $\hat{p}^2$ obtained from $B(x)$,
\begin{equation}
    \frac{1+K_Q}{2} = -\frac{1}{\hbar^2}\left(\mbox{Re}\left(\bra{\phi}\hat{x}^2\hat{p}^2\ket{\phi}\right) - \braket{\hat{x}^2}\braket{\hat{p}^2}\right).
\end{equation}
This lets us confirm the result previously obtained in \cite{MH24} using \Cref{eq:qvar},
\begin{equation}
    \Delta A_f^2 = \varepsilon_A^2(f) +  \left(\mbox{Re}\left(\bra{\phi}\hat{x}^2\hat{p}^2\ket{\phi}\right) - \braket{\hat{x}^2}\braket{\hat{p}^2}\right) \frac{\frac{\partial^2}{\partial \phi_A^2} P(f)}{2 P(f) }
\end{equation}
where the back action sensitivity of the post-selection probability $P(f)$ is given by Eq.(\ref{eq:sense}). 
In this decomposition, the Ozawa-Hall uncertainty $\varepsilon_A^2(f)$ represents the intrinsic fluctuations of the weak value $\braket{\hat{A}}_f$, while the term that depends on the dynamics of the post-selection probability represents a Bayesian update of the meter statistics. 

In the weak measurement regime, it is not possible to separate the meter statistics associated with diagonal elements of the density matrix from the interference terms $I(a,a',x)$ associated with quantum superpositions of different eigenvalues $A_a$. It is therefore necessary to consider the relation between the statistics expressed by the interference terms, and the two eigenvalues associated with them. By considering the second derivative in measurement strength, we have identified the precise effects of post-selection in a weak measurement on the interference patterns $I(a,a',x)$ of meter shifts associated with quantum superpositions of different eigenvalues $A_a$. Independent of meter state, we can identify a negative diffusion term that compensates the fluctuations of $\hat{A}$ that are introduced by separating the contributions associated with different eigenvalues from each other. We can conclude that such a separation is only meaningful when quantum interference between different eigenstates is negligible. The statistics of meter interference thus ensure that the statistics of weak measurements are fundamentally different from the statistics of eigenvalues. It has been pointed out that post-selection can result in statistical artefacts because statistical correlations between the post-selected outcome and the average meter shift could create the illusion of anomalous values of a system property \cite{FC14}. Our result shows why this criticism of post-selected measurements does not apply to the physics of weak measurements. The correlation between post-selection probability and meter readout exists, but it only impacts the post-selected meter statistics in second order in measurement strength, where it complicates the interpretation of conditional meter fluctuations.

\section{Readout statistics without Bayesian artefacts}
\label{sec:nonBayesian}

Above we showed that the normalised distribution of the symmetric meter interference $I(a,a',x)$ can be decomposed into two terms: a negative diffusion term, that originates from the post-selected statistics of the system observable $\hat{A}$; and a term that describes a Bayesian update, originating from the information about the magnitude of the squared meter momentum $\hat{p}^2$ from the observation of the post-selected outcome $f$. We stress that this modification of the readout distribution is not caused by changes in the meter position. It is therefore important to distinguish the signature of a meter update from the signature of the negative diffusion term that represents a more fundamental relation between meter interference and the quantum statistics of the system observable. 

As discussed in \cite{MH24}, Gaussian meter states are characterised by a negative correlation between squared position and squared momentum. For a Gaussian state $\ket{G}$ with variance $\sigma$, the Bayesian update is given by
\begin{equation}
    B_G(x) = \frac{1}{2 \sigma^4}\left(x^2-\sigma^2 \right) |\braket{x|G}|^2.
\end{equation}
The Bayesian update is negative for $|x|<\sigma$, and positive for $|x|>\sigma$. It is indistinguishable from a diffusive broadening of the Gaussian distribution, 
\begin{equation}
    B_G(x) = \frac{1}{2} \frac{\partial^2}{\partial x^2} |\braket{x|G}|^2.
\end{equation}
For Gaussian meter states, $I(a,a',x)=|\braket{x|G}|^2$ is the same at all measurement strengths $s$ because the negative diffusion and the Bayesian update compensate each other. The meter state characteristic $K_Q$ is zero, so the weak conditional uncertainty is given by the quasi-probability variance $V_Q$. The experimentally observed readout fluctuations of Gaussian meters thus appear to confirm the quasi-probability $Q(a,a'|f)$. However, this is a statistical artefact caused by the Bayesian update characteristics of Gaussian meter states and does not represent the quantum statistics of the post-selected system.
For most other meter wavefunctions, the Bayesian meter update can be distinguished from the diffusion term.

Let us consider how to avoid Bayesian update effects in the meter state statistics. This can be achieved by using a meter state with an update characteristic of $B(x)=0$ for all relevant values of $x$, so that the meter state $\ket{\phi_0}$ satisfies
\begin{equation}
\label{eq:Bzero}
\textrm{Re}\left(\frac{\bra{x}\hat{p}^2\ket{\phi_0}}{\braket{x|\phi_0}}\right)=\braket{\phi_0|\hat{p}^2|\phi_0}.
\end{equation}
The solutions that satisfy this condition for all relevant positions $x$ are equivalent to the energy eigenstates in an infinite potential well, where the boundary condition minimises the contribution of the discontinuities at the edges of the interval on which the wavefunction is defined. For a single maximum, the normalised wavefunction reads
\begin{equation}
    \braket{x|\phi_0} = \sqrt{\frac{2}{L}} \cos\left(\frac{\pi}{L} x\right)
    \hspace{0.3cm} \mbox{for} \hspace{0.3cm} x\leq L/2.
\end{equation}
The momentum uncertainty of this state is inversely proportional to the position uncertainty,
\begin{equation}
    \bra{\phi_0} \hat{p^2} \ket{\phi_0} = \left(\frac{\pi \hbar}{L} \right)^2.
\end{equation}
For these cosine-shaped meter states, $B(x)=0$ indicates the absence of Bayesian updates, and $K_Q$ has a value of $-1$. For sufficiently small meter shifts, the weak conditional uncertainty is equal to the Ozawa-Hall uncertainty $\varepsilon_A(f)$ of the post-selected system. The changes to the interference patterns are described by the negative diffusion term
\begin{equation}
    -\frac{1}{2} \frac{\partial^2}{\partial x^2} |\braket{x|\phi_0}|^2 = 2 \left(\frac{\pi}{L}\right)^2 \left(|\braket{x|\phi_0}|^2-\frac{1}{L} \right).
\end{equation}
The negative diffusion term subtracts a flat background term from the readout distribution, compensating the flat background generated by the sum over distributions with different meter shifts. For cosine-meter wavefunctions, fluctuations in $\hat{A}$ reduce the contrast of the meter distribution by adding a flat background. The approximate readout distribution takes the form
\begin{equation}
    P_{\phi_0}(x|f) = \frac{1}{L} \left(1 + \nu \cos\left(\frac{2 \pi}{L}(x-s \braket{\hat{A}}_f)\right) \right)
\end{equation}
with the visibility $\nu$ of the meter distribution given by
\begin{equation}
    \nu = 1-2 \left(\frac{\pi}{L} \right)^2 s^2 \varepsilon_A^2(f).
\end{equation}
It is therefore possible to determine the Ozawa-Hall uncertainty $\varepsilon_A^2(f)$ directly from the reduction of visibility associated with the addition of a diffusive background to the readout distribution of the cosine wavefunction of the meter.  

Note that the discontinuities of cosine meter wavefunctions at $x=\pm L/2$ make it difficult to evaluate experimental results close to these discontinuities. However, the discussion above shows that the main results can be obtained at a sufficient distance from the discontinuities of the wavefunction gradient. Cosine wavefunctions should therefore be considered as viable alternatives to the Gaussian meter wavefunctions commonly used. As shown above, Gaussian meter wavefunctions make it particularly difficult to separate the effects of Bayesian updates from the intrinsic fluctuations of meter shifts induced by the values of $\hat{A}$. When other meter wavefunctions are used, one can distinguish Bayesian updates from the diffusive change caused by fluctuating meter shifts. In all cases, it is important to recognise the effects that system post-selection may have on the statistics of the meter.

\section{The role of meter interference in quantum measurements}
\label{sec:meterinterference}

Meter interference is a phenomenon that can only be observed in post-selected measurements. It might therefore be tempting to dismiss it as tangential, of no consequence for a more general understanding of quantum statistics. However, post-selection is a legitimate procedure that provides additional information on a system and on its interactions with a meter. This additional information makes it impossible to interpret weak measurements in terms of the eigenvalues of the target observable $\hat{A}$. Attempts to construct classical statistical models to restore an overly realist interpretation of eigenvalues lack merit, since they completely ignore the physics of the measurement interaction \cite{FC14}. The mathematical description of measurement clearly identifies quantum interference effects as the origin of weak values outside of the spectrum of possible eigenvalues \cite{Dre15}. However, quantum interference itself has no clear physical meaning. We would be fooling ourselves if we claimed that we know what it means that the probability of observing $f$ is determined by interferences between two eigenvalues of an unobserved physical property. In a post-selected weak measurement, the quantum interference effects that determine the post-selection probability $P(f)$ also appear as interference effects in the conditional readout statistics of the meter system. Mathematically, it is easy to identify the contributions associated with quantum interference between different meter shifts, where the contributions to $P(f)$ define a quasi-probability $Q(a,a'|f)$ that represents destructive interference as a negative probability. As shown in \Cref{sec:model}, this quasi-probability identifies the real parts of the weak value as conditional averages of the quantum statistics described by the initial state of the system. However, the formalism indicates that interference patterns between different meter shifts differ from the initial meter distribution by more than just their displacement. An increase in measurement strength results in a gradual distortion of the meter statistics contributed by interferences between different meter shifts. 

In the present analysis, we have shown that this change of the interference pattern with increasing measurement strength can be traced back to the respective roles played by intrinsic uncertainties of $\hat{A}$, and by back-action effects on the post-selection probabilities $P(f)$. As measurement strength increases, post-selection inadvertently provides information on the magnitude of the meter momentum $\hat{p}$, since back-action effects are stronger for high values of $\hat{p}^2$. This information shows up in the interference patterns for different meter shifts as high values of $\hat{p}^2$ reduce the coherences between different eigenvalues of $\hat{A}$, modifying the contributions of quantum interference to the meter statistics. Here, the role of quantum coherence between different eigenvalues of $\hat{A}$ is dynamical - the back-action reduces the coherence by the momentum dependence of phase shifts, while diagonal elements of the density matrix remain unchanged. Since this modification of meter statistics arises from information about the meter momentum, it is unrelated to the values of $\hat{A}$ that determine the possible meter shifts.  

On the other hand, the negative diffusion term identified in \Cref{sec:analysis} does reveal something fundamental about the relation between quantum interference and the statistics of $\hat{A}$ conditioned by post-selection. This term modifies the fluctuations described by the quasi-probability $Q(a,a'|f)$, by lowering the variance associated with interference terms. In the case of negative $Q(a,a'|f)$, this results in an increase of fluctuations in the meter shift, which is important because negative probabilities can result in negative fluctuations. The negative diffusion term compensates all negative contributions in the quasi-probability variance $V_Q$, modifying the fluctuations so that they correspond to the Ozawa-Hall uncertainties $\varepsilon_A^2(f)$. Since this happens in a regime where different meter shifts overlap, the negative diffusion term ensures that the fluctuations associated with the conditional averages $\braket{\hat{A}}_f$ observed in weak measurements are different from the fluctuations required by a distribution of eigenvalues with that average. In the case of pure state inputs and real weak values, the negative diffusion term ensures that $\varepsilon_A^2(f)=0$, indicating that the weak value is the dispersion-free value of $\hat{A}$ conditioned by this set of quantum interferences. Meter interference thus modifies the set of individual values that determine the conditional meter shifts in a weak measurement. 

This might seem a bold conclusion. However, there is strong evidence in its favour. Measurement interactions cannot be described in terms of classical statistics since they generally entangle the system with the meter, even when the interaction is very weak. Post-selection erases all available information about eigenstates of the observable $\hat{A}$, leaving only a superposition of shifted meter states as a quantum-mechanical record of $\hat{A}$. We have shown that the interference between the different meter states modifies not only the conditional average, but also the conditional fluctuations. Meter interference thus modifies the statistics of conditional meter shifts in a way that restricts the set of possible values that the meter shifts can take, most strikingly assigning an uncertainty of zero to the real weak values of pure states. Given that there is no positive quasi-probability that could associate the post-selected outcome $f$ with the eigenvalues of $\hat{A}$, it seems feasible that the initial state uncertainty of the observable $\hat{A}$ can be resolved using a different set of conditional values of $\hat{A}$, depending on the final measurement context determined by the post-selection. At the least, the detailed analysis of meter interference given above should clarify the manner in which quantum effects modify the statistical evidence associated with meter shifts proportional to the observable $\hat{A}$. The results indicate that the complete statistics of a meter readout provide much more information about the physics of the system-meter interaction and its relation with the post-selected outcome $f$ than the study of conditional averages on which the debate has focused previously.

\section{Conclusions}
\label{sec:conclusions}

In this work, we have performed a detailed analysis of the effects of meter interference on the readout statistics of post-selected weak measurements. We started from the observation that the conditional averages corresponding to the real parts of the weak value can be explained using a quasi-probability defined by the contributions of different density matrix elements to the post-selection probability. We used this to consider the measurement strength dependence of the interference pattern between different meter shifts. The variance of this interference pattern changes with measurement strength at a rate that depends on the initial meter wavefunction. As two of us showed in previous work \cite{MH24}, this change in the conditional readout fluctuations includes the effects of correlations between the meter readout and the squared meter momentum, because it is the meter momentum that determines the effects of back-action on the post-selection probability. In the present work, we have shown that this effect is described in detail by the changes in the meter interference patterns $I(a,a',x)$. Post-selection causes a Bayesian update of the readout distribution in the initial state. On the other hand, the change of the interference pattern traceable to intrinsic fluctuations of the meter shift is described by a negative diffusion term. This negative diffusion term ensures that the observed variance in the meter shifts is given by Ozawa-Hall uncertainties, effectively removing the possibility of negative fluctuations associated with the quasi-probability interpretation used in the initial analysis. Any reduction of readout fluctuations can thus be traced to the Bayesian update of the meter statistics. For instance, the Gaussian meter squeezing interpreted as a weak variance in Ref.~\cite{Ogawa} is explained by the positive correlation of the post-selection condition $f$ with the high weak values of $\hat{p}^2$ found near the centre of the readout distribution. 

Our results show that quantum interference changes the meter readout fluctuations in two ways: a Bayesian update, and a negative diffusion term. The negative diffusion term serves as a correction of the quasi-probability statistics obtained by interpreting the contributions from different density matrix elements as separate events. This negative diffusion term suggests that the correct form of the conditional fluctuations of the observable $\hat{A}$ in a post-selected weak measurement are given by the Ozawa-Hall uncertainties $\varepsilon_A^2(f)$. Here, the negative diffusion term ensures that the fluctuations of real weak values are strongly suppressed, dropping all the way to zero in the case of pure states. Quantum interference thus characterises a deterministic relation between weak values and eigenvalues, by modifying the statistical noise of the meter shift distribution. We stress that this observation helps to explain how weak values can depend on the post-selected outcomes of a future measurement. Before such a future measurement occurs, the only empirical record of the quantity $\hat{A}$ available anywhere in the world is the data obtained from the meter readout. Since the meter itself is a quantum system, one should not assume that the uncertainty of the meter readout is only a technical or epistemic limitation. As our analysis shows, quantum interference in the meter ensures that weak values emerge as precise values of $\hat{A}$ in the quantum correlations between the meter statistics and the post-selected outcome, no matter what that outcome may be. Meter interference decides which set of values a physical property can have, exceeding any statistical limits set by its eigenvalues. This loss of certainty regarding the unconditional validity of eigenvalues has far-reaching consequences. As shown in recent experiments, the presence of a particle in a path is a quantity that can seemingly take values other than zero or one, allowing the quantitative characterisation of a physical delocalisation of the particle across several paths in an interferometer~\cite{Lem+22,Fukuda}. The analysis of meter interference presented above shows that such phenomena are part of a consistent description of measurement interactions, paving the way towards a more complete understanding of quantum physics as an empirical science.



\section*{Acknowledgments}
This work was supported by ERATO, Japan Science and Technology Agency (JPMJER2402). JRH acknowledges support from a Royal Society Research Grant (RG/R1/251590), an EPSRC Mathematical Sciences Small Grant (UKRI3647), and their EPSRC Quantum Technologies Career Acceleration Fellowship (UKRI1217).

\nocite{*}
\bibliography{ref.bib}

\end{document}